\documentclass[jmp,showpacs,onecolumn,amsmath,amsfonts,amssymb,superscriptaddress,nofootinbib,10pt]{revtex4}
\usepackage{graphicx}
\usepackage{amsmath}
\usepackage{float}
\usepackage{longtable}
\usepackage{epsfig}
\usepackage{amssymb}
\usepackage{amsfonts}
\usepackage{latexsym}
\usepackage{theorem}
\usepackage{bbm}
\usepackage{bm}
\usepackage{psfrag}
\usepackage{geometry}

\newtheorem{theo}{Theorem}
\newtheorem{lemma}[theo]{Lemma}
\newtheorem{propo}[theo]{Proposition}
\newtheorem{coro}[theo]{Corollary}

\newcommand{\ket}[1]{\ensuremath{|#1 \rangle}}
\newcommand{\bra}[1]{\ensuremath{\langle #1|}}

\newcommand{\kb}[1]{\ensuremath{| #1 \rangle \! \langle #1 |}}

\newcommand{\ip}[2]{\ensuremath{\langle #1 | #2 \rangle}}
\newcommand{\op}[2]{\ensuremath{| #1 \rangle \! \langle #2 |}}

\newcommand{\1}{\protect\ensuremath{\mathbbm{1}}}
\newcommand{\F}{\ensuremath{\mathbbm{F}}}

\newcommand{\abs}[1]{\ensuremath{| #1 |}}
\newcommand{\cb}[1]{\ensuremath{\| #1 \|_{cb}}}
\newcommand{\norm}[1]{\ensuremath{\| #1 \|_{\infty}}}
\newcommand{\tracenorm}[1]{\ensuremath{\| #1 \|_{1}}}

\newcommand{\C}{\ensuremath{\mathbb{C}}}

\newcommand{\N}{\ensuremath{\mathbb{N}}}

\newcommand{\tr}[1]{\ensuremath{{\rm tr}(#1)}}
\newcommand{\ld}{\ensuremath{{\rm ld} \, }}
\newcommand{\id}{\ensuremath{{\rm id} \, }}

\newcommand{\re}[1]{\ensuremath{{\rm Re}(#1)}}

\newcommand{\hh}{\ensuremath{\mathcal{H}}}
\newcommand{\bh}{\ensuremath{\mathcal{B(H)}}}
\newcommand{\bhh}[1]{\ensuremath{\mathcal{B}(\mathcal{H}_{#1})}}
\newcommand{\bc}[1]{\ensuremath{\mathcal{B}(\mathbb{C}^{#1})}}
\newcommand{\bcstar}[1]{\ensuremath{\mathcal{B_{*}}(\mathbb{C}^{#1})}}

\newcommand{\bhstar}{\ensuremath{\mathcal{B_{*}(H)}}}
\newcommand{\bhhstar}[1]{\ensuremath{\mathcal{B}_{*}(\mathcal{H}_{#1})}}

\newcommand{\A}{\ensuremath{\mathcal{A}}}
\newcommand{\B}{\ensuremath{\mathcal{B}}}

\begin{document}

\title{The Information-Disturbance Tradeoff\\ and the Continuity of
Stinespring's Representation}
\author{Dennis Kretschmann}
\email{d.kretschmann@tu-bs.de} \affiliation{Department of Applied
Mathematics and Theoretical Physics, University of Cambridge,
Wilberforce Road, Cambridge CB3 0WA, United Kingdom}
\affiliation{Institut f\"ur Mathematische Physik, Technische
Universit\"at Braunschweig, Mendelssohnstra{\ss}e~3, 38106
Braunschweig, Germany}
\author{Dirk Schlingemann}
\email{d.schlingemann@tu-bs.de} \affiliation{Institut f\"ur
Mathematische Physik, Technische Universit\"at Braunschweig,
Mendelssohnstra{\ss}e~3, 38106 Braunschweig, Germany}
\author{Reinhard F.~Werner}
\email{r.werner@tu-bs.de} \affiliation{Institut f\"ur
Mathematische Physik, Technische Universit\"at Braunschweig,
Mendelssohnstra{\ss}e~3, 38106 Braunschweig, Germany}

\date{30~April~2006}

\begin{abstract}
    Stinespring's dilation theorem is the basic structure theorem
    for quantum channels: it states that any quantum channel
    arises from a unitary evolution on a larger system. Here we
    prove a continuity theorem for Stinespring's dilation: if two
    quantum channels are close in cb-norm, then it is always
    possible to find unitary implementations which are close in
    operator norm, with dimension-independent bounds. This result
    generalizes Uhlmann's theorem from states to channels and
    allows to derive a formulation of the information-disturbance
    tradeoff in terms of quantum channels, as well as a continuity
    estimate for the no-broadcasting theorem. We briefly discuss
    further implications for quantum cryptography, thermalization
    processes, and the black hole information loss puzzle.
\end{abstract}


\maketitle

%%%%%%%%%%%%%%%%%%%%%%%%%%%%%%%%%%%%%%%%%%%%%%%%%%%%%%%%%%%%%%%%%%%%%%%%%%%%%%%%%%
%%%%%%%%%%%%%%%%%%%%%%%%%%%%%%%%%%%%%%%%%%%%%%%%%%%%%%%%%%%%%%%%%%%%%%%%%%%%%%%%%%
%%%%%%%%%%%%%%%%%%%%%%%%%%%%%%%%%%%%%%%%%%%%%%%%%%%%%%%%%%%%%%%%%%%%%%%%%%%%%%%%%%
%%%%%%%%%%%%%%%%%%%%%%%%%%%%%%%%%%%%%%%%%%%%%%%%%%%%%%%%%%%%%%%%%%%%%%%%%%%%%%%%%%

\section{Introduction}
    \label{sec:intro}

According to Stinespring's dilation theorem \cite{Sti55}, every
completely positive and trace-preserving map, or {\em quantum
channel}, can be built from the basic operations of $(i)$
tensoring the input with a second system in a specified state
(conventionally called the {\em ancilla system}), $(ii)$ unitary
transformation on the combined input -- ancilla system, and
$(iii)$ reduction to a subsystem. Any channel can hence be thought
of as arising from a unitary evolution on a larger ({\em dilated})
system. The theorem comes with a bound on the dimension of the
ancilla system. Stinespring's dilation thus not only provides a
neat characterization of the set of permissible quantum
operations, but is also a most useful tool in quantum information
science.

Our contribution is a continuity theorem for Stinespring's
dilation: we show that two quantum channels, $T_{1}$ and $T_{2}$,
are close in cb-norm iff we can find dilating unitaries, $V_{1}$
and $V_{2}$, that are close in operator norm:
\begin{equation}
        \label{eq:intro01}
            \inf_{V_{1},V_{2}} \, \norm{V_{1} - V_{2}}^2
            \leq \cb{T_{1} - T_{2}} \leq 2 \, \inf_{V_{1},V_{2}}
            \norm{V_{1} - V_{2}}.
\end{equation}
The cb-norm $\cb{\cdot}$ that appears in Eq.~(\ref{eq:intro01}) is
a stabilized version of the standard operator norm $\norm{\cdot}$,
as explained in Sec.~\ref{sec:distance}.

Stinespring's representation is unique up to unitary
transformations on the ancilla system. So we may just as well fix
two Stinespring dilations $V_{1}$ and $V_{2}$ for $T_{1}$ and
$T_{2}$, respectively, and optimize over all unitaries $U$ on the
ancilla system. The continuity estimate Eq.~(\ref{eq:intro01}) can
then be rewritten as
\begin{equation}
        \label{eq:intro02}
            \inf_{U} \, \norm{ (\1_{B} \otimes U) V_{1} - V_{2}}^2
            \leq \cb{T_{1} - T_{2}} \leq 2 \, \inf_{U}
            \norm{(\1_{B} \otimes U) V_{1} - V_{2}}
\end{equation}
(cf. Th.~\ref{theo:continuity} in Sec.~\ref{sec:continuity}).
Hence, the continuity theorem generalizes the uniqueness clause in
Stinespring's theorem to cases in which two channels $T_{1}$,
$T_{2}$ differ by a finite amount. For states, i.\,e., channels
with one-dimensional domain, dilations are usually called {\it
purifications}, and in this special case Eq.~(\ref{eq:intro02}) is
an immediate consequence of Uhlmann's theorem. The proof of the
continuity theorem relies on a generalization of Uhlmann's theorem
from quantum states to quantum channels, and will be presented in
Sec.~\ref{sec:theorem} --- preceded by a brief introduction to
quantum channels and distance measures in Sec.~\ref{sec:prelim}.
We initially restrict our discussion to finite-dimensional Hilbert
spaces. Yet the continuity estimate Eq.~(\ref{eq:intro02}) has the
welcome feature of being completely independent of the dimension
of the underlying Hilbert spaces, and is thus perfectly tailored
for applications in which this dimension is unknown or large. In
Sec.~\ref{sec:summary} we will briefly describe extensions of our
results to infinite-dimensional systems.

The ancilla system in Stinespring's representation has a natural
interpretation as the environment of the physical system under
investigation: the output of the channel $T$ arises from a unitary
interaction of the input state with the environment, followed by a
partial trace over the degrees of freedom of the environment. Any
channel $T_{}$ then has a {\em complementary channel} $T_{E}$, in
which the roles of the output system and the environment are
interchanged. $T_{E}$ describes the information flow into the
environment. Since complementary channels share a common
Stinespring representation, Eq.~(\ref{eq:intro02}) allows to
relate the distance between two quantum channels to the distance
between their complementaries. This is particularly fruitful for
the noiseless (or {\em ideal}) channel $\id$, whose complementary
channel $S$ is completely depolarizing. The continuity theorem
then entails a formulation of the information-disturbance
tradeoff, which lies at the heart of quantum physics and explains
why quantum information behaves so fundamentally different from
its classical counterpart. We prove in Sec.~\ref{sec:tradeoff}
that almost all the information can be retrieved from the output
of the quantum channel $T$ by means of a decoding operation $D$
iff $T$ releases almost no information to the environment:
\begin{equation}
        \label{eq:intro03}
            \frac{1}{4} \inf_{D} \cb{T D - \id}^2 \leq
            \cb{T_{E} - S_{}} \leq 2 \, \inf_{D} \cb{T D -
            \id}^{\frac{1}{2}}
\end{equation}
(cf. Th.~\ref{theo:tradeoff} in Sec.~\ref{sec:tradeoff}). Again,
no dimension-dependent factors appear in these bounds. However, we
show in Sec.~\ref{sec:alternative} that this welcome property
crucially depends on the choice of the operator topology: if the
cb-norm $\cb{\cdot}$ is replaced by the standard operator norm
$\norm{\cdot}$ in Eq.~(\ref{eq:intro03}), a dimension-independent
bound can in general no longer be given.

The tradeoff between information and disturbance guarantees the
security of quantum key distribution in a very strong form and
implies that quantum information cannot be cloned or distributed.
The tradeoff theorem then amounts to a continuity estimate for the
no-broadcasting theorem.

Further applications are briefly discussed in
Sec.~\ref{sec:further}, including thermalization processes and the
famous black hole information loss puzzle. These have been
investigated in more detail by Braunstein and Pati \cite{BP05}.
Finally, we show in a companion paper \cite{AKS+06} how the
continuity estimate allows to strengthen the impossibility proof
for quantum bit commitment.

%%%%%%%%%%%%%%%%%%%%%%%%%%%%%%%%%%%%%%%%%%%%%%%%%%%%%%%%%%%%%%%%%%%%%%%%%%%%%%%%%%
%%%%%%%%%%%%%%%%%%%%%%%%%%%%%%%%%%%%%%%%%%%%%%%%%%%%%%%%%%%%%%%%%%%%%%%%%%%%%%%%%%
%%%%%%%%%%%%%%%%%%%%%%%%%%%%%%%%%%%%%%%%%%%%%%%%%%%%%%%%%%%%%%%%%%%%%%%%%%%%%%%%%%
%%%%%%%%%%%%%%%%%%%%%%%%%%%%%%%%%%%%%%%%%%%%%%%%%%%%%%%%%%%%%%%%%%%%%%%%%%%%%%%%%%

\section{Preliminaries}
    \label{sec:prelim}

We begin with a brief introduction to quantum channels and
convenient measures to evaluate their distance. We refer to
Davies' textbook \cite{Dav76} and Keyl's survey article
\cite{Key02} for a more extensive discussion.

%%%%%%%%%%%%%%%%%%%%%%%%%%%%%%%%%%%%%%%%%%%%%%%%%%%%%%%%%%%%%%%%%%%%%%%%%%%%%%%%%%
%%%%%%%%%%%%%%%%%%%%%%%%%%%%%%%%%%%%%%%%%%%%%%%%%%%%%%%%%%%%%%%%%%%%%%%%%%%%%%%%%%

\subsection{Observables, States, and Channels}
    \label{sec:observables}

The statistical properties of quantum (as well as classical and
hybrid) systems are characterized by spaces of operators on a
Hilbert space $\hh$: The observables of the system are represented
by bounded linear operators on $\hh$, written $\bh$, while the
physical states associated with the system are the positive linear
functionals $\omega \mathpunct : \bh \rightarrow \C$ that satisfy
the normalization condition $\omega (\1) = 1$, with the identity
operator $\1 \in \bh$. We restrict this discussion to
finite-dimensional Hilbert spaces, for which all linear operators
are bounded and every linear functional $\omega$ can be expressed
in terms of a trace-class operator $\varrho_{\omega} \in \bhstar$
such that $\omega(a) = \tr{\varrho_{\omega} a}$ for all $a \in
\bh$. The normalization of the functional $\omega$ than translates
into the condition $\tr{\varrho_{\omega}} = 1$. In the
finite-dimensional setup, the physical states can thus be
identified with the set of normalized density operators $\varrho
\in \bhstar$. This correspondence no longer holds for
infinite-dimensional systems. Some of the necessary amendments
will be described in Sec.~\ref{sec:summary}.

A {\em quantum channel} $T$ which transforms input systems
described by a Hilbert space $\hh_{A}$ into output systems
described by a (possibly different) Hilbert space $\hh_{B}$ is
represented by a completely positive and unital map $T \mathpunct:
\bhh{B} \rightarrow \bhh{A}$. By unitality we mean that $T(\1_{B})
= \1_{A}$, with the identity operator $\1_{X} \in \bhh{X}$.
Complete positivity means that $\id_{\nu} \otimes T$ is positive
for all $\nu \in \N$, where $\id_{\nu}$ denotes the identity
operation on the $(\nu \times \nu)$ matrices.

The physical interpretation of the quantum channel $T$ is the
following: when the system is initially in the state $\varrho \in
\bhhstar{A}$, the expectation value of the measurement of the
observable $b \in \bhh{B}$ at the output side of the channel is
given in terms of $T$ by $\tr{\varrho \: T(b)}$.

Alternatively, we can focus on the dynamics of the states and
introduce the dual map $T_{*} \mathpunct : \bhhstar{A} \rightarrow
\bhhstar{B}$ by means of the duality relation
\begin{equation}
    \label{eq:dual}
        \tr{T_{*}(\varrho) \, b} = \tr{\varrho \, T(b)} \quad
        \forall \; \; \varrho \in \bhhstar{A}, b \in \bhh{B}.
\end{equation}
$T_{*}$ is a completely positive and trace-preserving map and
represents the channel in {\em Schr\"odinger picture}, while $T$
provides the {\em Heisenberg picture} representation. For
finite-dimensional systems, the Schr\"odinger and the Heisenberg
picture provide a completely equivalent description of physical
processes. The interconversion is always immediate from
Eq.~(\ref{eq:dual}).

%%%%%%%%%%%%%%%%%%%%%%%%%%%%%%%%%%%%%%%%%%%%%%%%%%%%%%%%%%%%%%%%%%%%%%%%%%%%%%%%%%
%%%%%%%%%%%%%%%%%%%%%%%%%%%%%%%%%%%%%%%%%%%%%%%%%%%%%%%%%%%%%%%%%%%%%%%%%%%%%%%%%%

\subsection{Distance Measures}
    \label{sec:distance}

For both the continuity theorem and the tradeoff theorem we will
need to evaluate the distance between different quantum channels
on the one hand, and different Stinespring isometries on the
other. There are several candidates for such distance measures,
which are adapted to different scenarios.

Assume two quantum channels, $T_{1}$ and $T_{2}$, with common
input and output spaces, $\hh_{A}$ and $\hh_{B}$, respectively.
Since these $T_{i}$ are (in Heisenberg picture) operators between
normed spaces $\bhh{B}$ and $\bhh{A}$, the natural choice to
quantify their distance is the {\em operator norm},
\begin{equation}
    \label{eq:distance01}
        \norm{T_{1} - T_{2}} := \sup_{b \neq 0} \frac{\norm{T_{1}(b) -
        T_{2}(b)}}{\norm{b}}.
\end{equation}
The norm distance Eq.~(\ref{eq:distance01}) has a neat operational
characterization: it is just twice the largest difference between
the overall probabilities in two statistical quantum experiments
differing only in replacing one use of $T_{1}$ with one use of
$T_{2}$.

However, in many applications it is more appropriate to allow for
more general experiments, in which the two channels are only
applied to a sub-system of a larger system. This requires {\em
stabilized} distance measures \cite{GLN04}, and naturally leads to
the so-called {\em norm of complete boundedness} (or {\em
cb-norm}, for short) \cite{Pau02}:
\begin{equation}
    \label{eq:distance02}
        \cb{T_{1} - T_{2}} := \sup_{\nu \in \N} \norm{ \id_{\nu} \otimes ( T_{1}
        - T_{2} )},
\end{equation}
where $\id_{\nu}$ again denotes the {\em ideal} (or {\em
noiseless}) channel on the $\nu$-dimensional Hilbert space
$\C^{\nu}$. Useful properties of the cb-norm include {\em
multiplicativity}, i.\,e., $\cb{T_{1} \otimes T_{2}} = \cb{T_1} \,
\cb{T_2}$, and {\em unitality}, $\cb{T} = 1$ for any channel
$T$.\\

Obviously, $\cb{T} \geq \norm{T}$ for every linear map $T$. If
either the input or output space is a classical system, we even
have equality: $\cb{T} = \norm{T}$ (cf. Ch.~3 in \cite{Pau02}).
Fully quantum systems generically show a separation between these
two norms. However, in the vicinity of the noiseless channel $\id$
the operator norm and the cb-norm may always be estimated in terms
of each other with dimension-independent bounds \cite{KW04}, and
can thus be considered equivalent, even when the dimensions of the
underlying Hilbert spaces are not known and possibly large:
\begin{equation}
    \label{eq:distance03}
        \norm{T-\id} \leq \cb{T-\id} \leq 8 \,
        \norm{T-\id}^{\frac{1}{4}}.
\end{equation}
Examples which show that this equivalence does {\em not} hold
generally will be provided in Sec.~{\ref{sec:alternative}. Thus,
in a quantum world correlations may help to distinguish locally
akin quantum channels. This plays an important role for the
interpretation of the tradeoff
theorem in Sec.~\ref{sec:tradeoff}.\\

States are channels with one-dimensional input space, $\hh_A =
\C$. Since this is a classical system, there is no need to
distinguish between stabilized and non-stabilized distance
measures. The so-called {\em trace norm} $\tracenorm{\varrho} =
{\rm tr \,} \sqrt{\varrho^{*} \varrho^{}}$ is a convenient measure
for the distance between two density operators. The trace norm
difference $\tracenorm{\varrho - \sigma}$ is equivalent to the
{\em fidelity} $f(\varrho,\sigma) := {\rm tr} \,
\sqrt{\sqrt{\varrho} \, \sigma \sqrt{\varrho}}$ by means of the
relation \cite{NC00}
\begin{equation}
    \label{eq:distance04}
        1 - f(\varrho,\sigma) \leq \frac{1}{2} \,
        \tracenorm{\varrho - \sigma} \leq \sqrt{1 - f^2(\varrho,
        \sigma)}.
\end{equation}

Finally, we note that for any linear operator $T$ the operator
norm $\norm{T}$ equals the norm of the Schr\"odinger adjoint
$T_{*}$ on the space of trace class operators, i.\,e.,
\begin{equation}
    \label{eq:distance05}
        \norm{T_{}} = \sup_{\tracenorm{\varrho} \leq 1} \;
        \tracenorm{T_{*} (\varrho)}
\end{equation}
(cf. Ch.~VI of \cite{RS80} and Sec.~2.4 of \cite{BR87} for
details), which is the usual way to convert norm estimates from
the Heisenberg picture into the Schr\"odinger picture and vice
versa. For states $T_{*} = \varrho$, the operator norm then indeed
just coincides with the trace norm: $\norm{T} = \tracenorm{T_{*}}
= \tracenorm{\varrho}$.

%%%%%%%%%%%%%%%%%%%%%%%%%%%%%%%%%%%%%%%%%%%%%%%%%%%%%%%%%%%%%%%%%%%%%%%%%%%%%%%%%%
%%%%%%%%%%%%%%%%%%%%%%%%%%%%%%%%%%%%%%%%%%%%%%%%%%%%%%%%%%%%%%%%%%%%%%%%%%%%%%%%%%
%%%%%%%%%%%%%%%%%%%%%%%%%%%%%%%%%%%%%%%%%%%%%%%%%%%%%%%%%%%%%%%%%%%%%%%%%%%%%%%%%%
%%%%%%%%%%%%%%%%%%%%%%%%%%%%%%%%%%%%%%%%%%%%%%%%%%%%%%%%%%%%%%%%%%%%%%%%%%%%%%%%%%

\section{Continuity of Stinespring's Representation}
    \label{sec:theorem}

\subsection{Stinespring's Representation}
    \label{sec:stinespring}

Stinespring's famous representation theorem \cite{Sti55,Pau02}, as
adapted to maps between finite-dimensional quantum systems, states
that any completely positive (not necessarily unital) map $T
\mathpunct: \bhh{B} \rightarrow \bhh{A}$ can be written as
\begin{equation}
    \label{eq:stinespring01}
        T(b) = V^{*} \left ( b \otimes \1_{E} \right ) V^{}
        \quad \forall \; \; b \in \bhh{B}
\end{equation}
with a linear operator $V \mathpunct : \hh_{A} \rightarrow \hh_{B}
\otimes \hh_{E}$. The finite-dimensional Hilbert space $\hh_{E}$
is usually called the {\em dilation space}, and the pair
$(\hh_{E}, V_{})$ a {\em Stinespring representation} for $T$. If
$T$ is unital (and thus a quantum channel), then $V$ is an {\em
isometry}, i.\,e., $V^{*} V^{} = \1_{A}$.

By means of the duality relation Eq.~(\ref{eq:dual}), in the
Schr\"odinger picture Stinespring's theorem gives rise to the
so-called {\em ancilla representation} of the quantum channel
$T_{*}$:
\begin{equation}
    \label{eq:stinespring02}
        T_{*} (\varrho) = {\rm tr }_{E} \, V^{} \varrho V^{*}
        \quad \forall \; \; \varrho \in \bhhstar{A},
\end{equation}
where ${\rm tr}_{E}$ denotes the {\em partial trace} over the
system $\hh_{E}$ \cite{NC00}. In the physical interpretation of
Stinespring's theorem the dilation space $\hh_{E}$ represents the
{\em environment}. Stinespring's isometry $V$ transforms the input
state $\varrho$ into the state $V^{} \varrho V^{*}$ on $\hh_{B}
\otimes \hh_{E}$, which is correlated between the output and the
environment. The output state $T_{*}(\varrho) \in \bhhstar{B}$ is
then obtained by tracing out the degrees of freedom of the
environment.  Physically, one would expect a unitary operation $U$
instead of an isometric $V$. However, the initial state of the
environment can be considered fixed, effectively reducing $U$ to
an isometry, $V \psi := U( \psi \otimes \psi_{0})$ for some fixed
initial pure state $\ket{\psi_{0}}$ of the environment system.

The Stinespring representation $(\hh_{E}, V_{})$ is called {\em
minimal} iff the set of vectors $( b \otimes \1_{E}) \, V \varphi$
with $b \in \bhh{B}$ and $\varphi \in \hh_{A}$ spans $\hh_{B}
\otimes \hh_{E}$. In this case the dilation space $\hh_{E}$ can be
chosen such that $\dim \hh_{E} \leq \dim \hh_{A} \, \dim \hh_{B}$.

%%%%%%%%%%%%%%%%%%%%%%%%%%%%%%%%%%%%%%%%%%%%%%%%%%%%%%%%%%%%%%%%%%%%%%%%%%%%%%%%%%
%%%%%%%%%%%%%%%%%%%%%%%%%%%%%%%%%%%%%%%%%%%%%%%%%%%%%%%%%%%%%%%%%%%%%%%%%%%%%%%%%%

\subsection{Uniqueness and Unitary Equivalence}
    \label{sec:unique}

Stinespring's representation is not at all unique: if $(\hh_{E},
V)$ is a representation for the completely positive map $T
\mathpunct: \bhh{B} \rightarrow \bhh{A}$, then it is easily seen
that a further representation for $T$ is given by $(\hh_{E},
(\1_{B} \otimes U) V)$ with any unitary $U \in \bhh{E}$.

However, it is straightforward to show that the minimal
Stinespring representation is unique up to such unitary
equivalence: Assume that the completely positive map $T
\mathpunct: \bhh{B} \rightarrow \bhh{A}$ has a minimal Stinespring
dilation $(\hh_{E}, V)$ as in Eq.~(\ref{eq:stinespring01}) as well
as a further, not necessarily minimal one $(\hh_{\widetilde{E}},
\widetilde{V})$, i.\,e.,
\begin{equation}
    \label{eq:unique01}
        T(b) = \widetilde{V}^{*} \left ( b \otimes \1_{\widetilde{E}} \right )
        \widetilde{V}^{} \quad \forall \; \; b \in \bhh{B}
\end{equation}
with another linear map $\widetilde{V} \mathpunct : \hh_{A}
\rightarrow \hh_{B} \otimes \hh_{\widetilde{E}}$ and a possibly
different dilation space $\hh_{\widetilde{E}}$. Since the
representation $(\hh_{E}, V)$ is chosen to be minimal, we conclude
that $\dim \hh_{E} \leq \dim \hh_{\widetilde{E}}$. Setting
\begin{equation}
    \label{eq:unique02}
        \widetilde{U} (b \otimes \1_{E}) V \varphi := (b \otimes
        \1_{\widetilde{E}}) \widetilde{V} \varphi
\end{equation}
for $b \in \bhh{B}$ and $\varphi \in \hh_{A}$ then yields a
well-defined isometry $\widetilde{U} \mathpunct : \hh_{B} \otimes
\hh_{E} \rightarrow \hh_{B} \otimes \hh_{\widetilde{E}}$. In
particular, by choosing $b = \1_{B}$ in Eq.~(\ref{eq:unique02}) we
see that $\widetilde{U} V = \widetilde{V}$. From the definition of
$\widetilde{U}$ we immediately find the intertwining relation
\begin{equation}
    \label{eq:unique03}
        \widetilde{U} (b \otimes \1_{E}) = (b \otimes \1_{\widetilde{E}})
        \widetilde{U} \quad \forall \; \; b \in \bhh{B}.
\end{equation}
Hence $\widetilde{U}$ must be decomposable as $\widetilde{U} =
\1_{B} \otimes U$ for some isometry $U \mathpunct : \hh_{E}
\rightarrow \hh_{\widetilde{E}}$. If both representations
$(\hh_{E}, V)$ and $(\hh_{\widetilde{E}}, \widetilde{V})$ are
minimal, the dimensions of the dilation spaces $\hh_{E}$ and
$\hh_{\widetilde{E}}$ coincide, and $U$ is unitary, as suggested.

%%%%%%%%%%%%%%%%%%%%%%%%%%%%%%%%%%%%%%%%%%%%%%%%%%%%%%%%%%%%%%%%%%%%%%%%%%%%%%%%%%
%%%%%%%%%%%%%%%%%%%%%%%%%%%%%%%%%%%%%%%%%%%%%%%%%%%%%%%%%%%%%%%%%%%%%%%%%%%%%%%%%%

\subsection{A Continuity Theorem}
    \label{sec:continuity}

We have seen from Stinespring's theorem that two minimal dilations
of a given quantum channel are unitarily equivalent. The
uniqueness clause is a powerful tool and has proved helpful in the
investigation of localizable quantum channels \cite{ESW01} and the
structure theorem for quantum memory channels \cite{KW05}. We will
now generalize the uniqueness clause to cases in which two quantum
channels differ by a finite amount, and prove continuity: if two
quantum channels are close in cb-norm, we may find Stinespring
isometries that are close in operator norm.\\

The converse also holds, and is in fact much simpler to show. So
we start with this part: Assume two quantum channels $T_{1}, T_{2}
\mathpunct : \bhh{B} \rightarrow \bhh{A}$ with Stinespring
isometries $V_{1}, V_{2} \mathpunct : \hh_{A} \rightarrow \hh_{B}
\otimes \hh_{E}$. We can always assume that $T_1$ and $T_2$ share
a common dilation space $\hh_{E}$, possibly after adding some
extra dimensions to one of the dilation spaces and some unitary
transformations. We do not assume that either dilation $(\hh_{E},
V_1 )$ or $(\hh_{E}, V_2 )$ is minimal.

A straightforward application of the triangle inequality shows
that for all $X \in \mathcal{B} (\C^{\nu}) \otimes \bhh{B}$ we
have
\begin{equation}
    \label{eq:continuity01}
        \begin{split}
            \norm{ \big ( \id_{\nu} \, \otimes & \, (T_{1} - T_{2}) \big ) \, X} \\
            & =  \norm{(\1_{\nu} \otimes V_{1}^{*}) X (\1_{\nu} \otimes V_{1}^{})
            \, - \, (\1_{\nu} \otimes V_{2}^{*}) X (\1_{\nu} \otimes V_{2}^{})}\\
            & \leq \norm{ \big [ \1_{\nu} \otimes (V_{1}^{*} - V_{2}^{*})
            \big ] \, X \, (\1_{\nu} \otimes V_{1}^{})}  +
            \, \norm{(\1_{\nu} \otimes V_{2}^{*}) \, X \, \big [
            \1_{\nu} \otimes (V_{1}^{} - V_{2}^{}) \big ]}\\
            & \leq 2 \, \norm{V_{1} - V_{2}} \, \norm{X},
        \end{split}
\end{equation}
independently of $\nu \in \N$, which immediately implies that
\begin{equation}
    \label{eq:continuity02}
        \cb{T_{1} - T_{2}} \leq 2 \, \norm{V_{1} - V_{2}}.
\end{equation}
Thus, if we can find Stinespring isometries $V_1$ and $V_2$ for
the channels $T_{1}$ and $T_{2}$ which are close in operator norm,
the channels will be close in cb-norm (and hence also in operator
norm, cf. Sec.~\ref{sec:distance}).\\

As advertised, we will now show the converse implication. As
Stinespring isometries are unique only up to unitary equivalence,
we cannot expect that {\em any} two given Stinespring isometries
$V_{1}, V_{2}$ are close. The best we can hope for is that these
isometries can be {\em chosen} close together, with
dimension-independent bounds. This is the essence of the
continuity theorem:
%%%%%%%%%%%%%%%%%%%%%%%%%%%%%%%%%%%%%%%%%%%%%%%%%%%%%%%%%%%%%%%%%%%%%%%%%%%%%%%%%%
\begin{theo}
    \label{theo:continuity}
    {\bf (Continuity)}\\
    Let $\hh_{A}$ and $\hh_{B}$ be finite-dimensional Hilbert
    spaces, and suppose that
    \begin{equation}
        \label{eq:continuity03}
            T_{1}, T_{2} \mathpunct : \bhh{B} \rightarrow
            \bhh{A}
    \end{equation}
    are quantum channels with Stinespring isometries
    $V_{1}, V_{2} \mathpunct : \hh_{A} \rightarrow \hh_{B} \otimes
    \hh_{E}$ and a common dilation space $\hh_{E}$. We then have:
    \begin{equation}
        \label{eq:continuity04}
            \inf_{U} \, \norm{ (\1_{B} \otimes U) V_{1} - V_{2}}^2
            \leq \cb{T_{1} - T_{2}} \leq 2 \, \inf_{U}
            \norm{(\1_{B} \otimes U) V_{1} - V_{2}},
    \end{equation}
    where the minimization is with respect to all unitary $U \in
    \bhh{E}$.
\end{theo}
%%%%%%%%%%%%%%%%%%%%%%%%%%%%%%%%%%%%%%%%%%%%%%%%%%%%%%%%%%%%%%%%%%%%%%%%%%%%%%%%%%

As a first step towards the proof of Th.~\ref{theo:continuity} we
will lift the equivalence Eq.~(\ref{eq:distance04}) of fidelity
and trace norm from quantum states to quantum channels. The
stabilized version of the fidelity for two quantum channels
$T_{1}$, $T_{2}$ has been called {\em operational fidelity}
\cite{BAR05}:
\begin{equation}
    \label{eq:continuity05}
        \begin{split}
            F(T_{1},T_{2}) :\!&= \inf \, \Big \{ f(\id \otimes T_{1*}
            \, \varrho, \, \id \otimes T_{2*} \, \varrho) \;
            \mid \; \varrho \in \bhhstar{A}^{\otimes 2},
            \tracenorm{\varrho} \leq 1 \Big \} \\
            & = \inf \, \Big \{ f(\id \otimes T_{1*}
            \, \kb{\psi}, \, \id \otimes T_{2*} \, \kb{\psi}) \;
            \mid \; \psi \in \hh_{A}^{\otimes 2}, \| \psi \|
            \leq 1 \Big \},
        \end{split}
\end{equation}
where minimization over pure states is sufficient by the joint
concavity of the fidelity $f$ (cf. \cite{NC00}, Th.~9.7).\\

Since quantum states are quantum channels with one-dimensional
domain (and stabilization is not needed in this case), we have
$F(\varrho, \sigma) = f(\varrho,\sigma)$ for any two quantum
states $\varrho, \sigma \in \bhhstar{A}$. The following Lemma,
which we again cite from \cite{BAR05}, is then a straightforward
generalization of the equivalence relation
Eq.~(\ref{eq:distance04}):
%%%%%%%%%%%%%%%%%%%%%%%%%%%%%%%%%%%%%%%%%%%%%%%%%%%%%%%%%%%%%%%%%%%%%%%%%%%%%%%%%%
\begin{lemma}
    \label{lemma:equivalence}
        {\bf (Equivalence)}\\
        For any two quantum channels $T_{1}, T_{2} \mathpunct : \bhh{B} \rightarrow
        \bhh{A}$ we have:
        \begin{equation}
            \label{eq:continuity06}
                1 - F(T_1, T_2) \leq \frac{1}{2} \, \cb{T_1 - T_2}
                \leq \sqrt{1 - F^2(T_1,T_2)},
        \end{equation}
        where $F(T_1,T_2)$ denotes the {\em operational fidelity}
        introduced in Eq.~(\ref{eq:continuity05}).
\end{lemma}
%%%%%%%%%%%%%%%%%%%%%%%%%%%%%%%%%%%%%%%%%%%%%%%%%%%%%%%%%%%%%%%%%%%%%%%%%%%%%%%%%%
{\bf Proof of Lemma~\ref{lemma:equivalence}}: The channel
difference $\Phi := T_1 - T_2$ is a linear map into the $\dim
\hh_{A}$-dimensional system $\bhh{A}$. Note that for such linear
maps $\Phi \mathpunct : \B \rightarrow \B (\C^{\nu})$,
stabilization with a $\nu$-dimensional bystander system is
sufficient, $\cb{\Phi} = \norm{\id_{\nu} \otimes \Phi}$ (cf.
\cite{Pau02}, Prop.~8.11). Conversion into the Schr\"odinger
picture via the duality relation Eq.~(\ref{eq:distance05}) then
yields
\begin{equation}
    \label{eq:continuity08}
        \cb{T_{1} - T_{2}} = \sup \, \Big \{
        \tracenorm{ \id \otimes (T_{1*} - T_{2*}) \, \varrho} \;
        \mid \; \varrho \in \bhhstar{A}^{\otimes 2},
        \tracenorm{\varrho} \leq 1 \Big \}.
\end{equation}
The statement of the lemma now immediately follows by combining
Eqs.~(\ref{eq:continuity05}) and (\ref{eq:continuity08}) with the
equivalence relation Eq.~(\ref{eq:distance04}). $\blacksquare$\\
%%%%%%%%%%%%%%%%%%%%%%%%%%%%%%%%%%%%%%%%%%%%%%%%%%%%%%%%%%%%%%%%%%%%%%%%%%%%%%%%%%

Lemma~\ref{lemma:equivalence} allows us to
concentrate entirely on fidelity estimates in the\\
\\
{\bf Proof of Th.~\ref{theo:continuity}:} According to Uhlmann's
theorem \cite{Uhl76,NC00}, the fidelity $f(\varrho,\sigma)$ of two
quantum states $\varrho, \sigma \in \bhhstar{A}$ has a natural
interpretation as the maximal overlap of all purifying vectors
$\psi_{\varrho}$, $\psi_{\sigma} \in \hh_{A} \otimes \hh_{R}$,
with the purification or reference system $\hh_{R} \cong \hh_{A}$:
\begin{equation}
    \label{eq:continuity09}
        f(\varrho,\sigma) = \max_{\psi_{\varrho}, \psi_{\sigma}} \,
        \abs{\ip{\psi_{\varrho}}{\psi_{\sigma}}}.
\end{equation}
In particular, it is possible to fix one of the purifications in
Eq.~(\ref{eq:continuity09}), $\psi_{\varrho}$ say, and maximize
over the purifications of $\sigma$. Since any two purifications of
the given state $\sigma$ are identical up to a unitary rotation $U
\in \bhh{R}$ on the purifying system, Uhlmann's theorem can be
given the alternative formulation
\begin{equation}
    \label{eq:continuity10}
        f(\varrho, \sigma) = \max_{U \in \bhh{R}} \,
        \abs{\ip{\psi_{\varrho}}{(\1_{A} \otimes U)
        \psi_{\sigma}}},
\end{equation}
where now $\psi_{\varrho}$ and $\psi_{\sigma}$ are any two {\em
fixed} purifications of $\varrho \in \bhhstar{A}$ and $\sigma \in
\bhhstar{A}$, respectively.

Since $(\1_{A'} \otimes V_{i}) \psi$ is a purification of the
output state $\id_{A'} \otimes T_{i*} (\kb{\psi})$, the
operational fidelity $F(T_1, T_2)$ can then be expressed in terms
of the Stinespring isometries $V_1$ and $V_2$ as follows:
\begin{equation}
    \label{eq:continuity11}
        \begin{split}
            F(T_1,T_2) & = \inf_{\psi} \, f(\id_{A'} \otimes T_{1*} \,
            \kb{\psi}, \, \id_{A'} \otimes T_{2*} \,
            \kb{\psi})\\
            & = \inf_{\psi} \sup_{U} \abs{\ip{(\1_{A'} \otimes V_{1})
            \psi}{(\1_{A'} \otimes \1_{B} \otimes U) \, (\1_{A'} \otimes
            V_{2}) \psi}}\\
            & = \inf_{\varrho \in \bhhstar{A}} \sup_{U} \,
            \abs{{\rm tr} \, \varrho \, V_{1}^{*} (\1_{B} \otimes
            U_{}^{})V_{2}^{} \, }\\
            & = \inf_{\varrho \in \bhhstar{A}} \sup_{U}\; \re{{\rm tr} \,
            \varrho \, V_{1}^{*} (\1_{B} \otimes
            U_{}^{})V_{2}^{}},
        \end{split}
\end{equation}
where maximization is over all unitary $U \in \bhh{E}$.

This representation is almost what we need for the desired norm
estimate, since only the (fixed) Stinespring isometries $V_{1}$,
$V_{2}$ and the unitary operations $U$ on the ancilla system
appear. However, from the order in which the optimization in
Eq.~(\ref{eq:continuity11}) is performed it is clear that the
optimal unitary $U$ for the inner maximization will in general
depend on the quantum state $\varrho$, $U = U(\varrho)$. In order
to obtain a universal unitary, observe that for fixed $\varrho \in
\bhhstar{A}$ the inner variation can be written as $\sup_{U} \,
\abs{{\rm tr} \, X U}$ with $X := {\rm tr}_{B} \, V_{2}^{} \varrho
V_{1}^{*} \in \bhh{E}$. It is easily seen that this supremum is
attained when $U$ is the unitary from the polar decomposition
\cite{NC00} of $X$, and equals $\tracenorm{X}$. However, since
$\abs{{\rm tr} \, X Y} \leq \tracenorm{X} \norm{Y}$ for all $Y \in
\bhh{E}$, we can replace the supremum over all unitaries in
Eq.~(\ref{eq:continuity11}) by a supremum over all $U \in \bhh{E}$
such that $\norm{U} \leq 1$.

With this modification both variations in
Eq.~(\ref{eq:continuity11}) range over convex sets, and the target
function is linear is both inputs. Von Neumann's minimax theorem
\cite{Neu28,Sim98} then allows us to interchange the infimum and
supremum to obtain:
\begin{equation}
    \label{eq:continuity12}
        F(T_1, T_2) = \sup_{\norm{U} \leq 1} \inf_{\varrho \in \bhhstar{A}}
        \; \re{{\rm tr} \, \varrho \, V_{1}^{*} (\1_{B} \otimes
        U_{}^{})V_{2}^{}}.
\end{equation}
The optimization now yields a universal $U \in \bhh{E}$. In
addition we know from our discussion above that $U$ can always be
chosen to be unitary in Eq.~(\ref{eq:continuity12}). Since
$\norm{Y} = \sup_{\tracenorm{\varrho} \leq 1} \abs{{\rm tr} \,
\varrho Y}$ for any $Y \in \bhh{A}$, we may now conclude that
\begin{equation}
    \label{eq:continuity16}
        \begin{split}
            \inf_{U} \, \norm{(\1_{B} \otimes U) V_{1} - V_{2}}^2
            & = \inf_{U} \, \norm{\big ( V_{1}^{*} (\1_{B} \otimes U_{}^{*} -
            V_{2}^{*}) \big ) \, \big ( (\1_{B} \otimes U) V_{1}^{} -
            V_{2}^{} \big)}\\
            & = \inf_{U} \, \sup_{\varrho} \, {\rm tr} \, \varrho \,
            \big ( V_{1}^{*} (\1_{B} \otimes U_{}^{*}) -
            V_{2}^{*} \big ) \, \big ( (\1_{B} \otimes U) V_{1}^{} -
            V_{2}^{} \big)\\
            & = 2 - 2 \sup_{U} \,\inf_{\varrho} \re{{\rm tr} \, \varrho
            V_{1}^{*} (\1_{B} \otimes U) V_{2}^{}} \\
            & = 2 \, \big ( 1 - F(T_{1}, T_{2}) \big )\\
            & \leq \cb{T_{1} - T_{2}},
        \end{split}
\end{equation}
where in the last step we have applied
Lemma~\ref{lemma:equivalence}. This proves the left half of
Eq.~(\ref{eq:continuity04}). $\blacktriangle$

The right half, which we have seen is the easier part, follows
immediately from our discussion leading to
Eq.~(\ref{eq:continuity02}) above. Alternatively, one could apply
the right half of the equivalence lemma
Eq.~(\ref{eq:continuity06}) to obtain that
\begin{equation}
    \label{eq:continuity17}
        \cb{T_{1} - T_{2}} \leq 2 \, \sqrt{1 - F^{2}(T_1, T_2)} \leq 2
        \, \sqrt{2} \, \sqrt{1 - F(T_1, T_2)}.
\end{equation}
Note that without any need to invoke the minimax theorem, we can
now directly conclude from Eq.~(\ref{eq:continuity11}) that
\begin{equation}
    \label{eq:continuity18}
        1 - F(T_1, T_2) \leq  1 - \sup_{U} \, \inf_{\varrho} \,
        {\rm Re} \, {\rm tr} \, \varrho \, V_{1}^{*} (\1_{B} \otimes U) V_{2}^{}
        = \frac{1}{2} \inf_{U} \, \norm{(\1_{B} \otimes U) V_{1} -
        V_{2}}^{2}.
\end{equation}
Substituting Eq.~(\ref{eq:continuity18}) into
Eq.~(\ref{eq:continuity17}), we then find
\begin{equation}
    \label{eq:continuity19}
        \cb{T_{1} - T_{2}} \leq 2 \inf_{U} \, \norm{(\1_{B} \otimes U)
        V_{1} - V_{2}},
\end{equation}
and so we have in fact rediscovered the familiar upper bound on the cb-norm
distance. $\blacksquare$

%%%%%%%%%%%%%%%%%%%%%%%%%%%%%%%%%%%%%%%%%%%%%%%%%%%%%%%%%%%%%%%%%%%%%%%%%%%%%%%%%%
%%%%%%%%%%%%%%%%%%%%%%%%%%%%%%%%%%%%%%%%%%%%%%%%%%%%%%%%%%%%%%%%%%%%%%%%%%%%%%%%%%
%%%%%%%%%%%%%%%%%%%%%%%%%%%%%%%%%%%%%%%%%%%%%%%%%%%%%%%%%%%%%%%%%%%%%%%%%%%%%%%%%%
%%%%%%%%%%%%%%%%%%%%%%%%%%%%%%%%%%%%%%%%%%%%%%%%%%%%%%%%%%%%%%%%%%%%%%%%%%%%%%%%%%

\section{Information-Disturbance Tradeoff}
    \label{sec:tradeoff}

Due to the essential uniqueness of the Stinespring dilation
$(\hh_{E}, V)$, to every quantum channel $T \mathpunct: \bhh{B}
\rightarrow \bhh{A}$ we may associate a {\em complementary
channel} $T_{E} \mathpunct: \bhh{E} \rightarrow \bhh{A}$, in which
the roles of the output system $\hh_{B}$ and the environment
system $\hh_{E}$ are interchanged:
\begin{equation}
    \label{eq:tradeoff01}
        T_{E}(e) := V^{*} \left ( \1_{B} \otimes e \right ) V^{}
        \quad \forall \; \; e \in \bhh{E}.
\end{equation}
The channel $T_{E}$ describes the information flow from the input
system $\hh_{A}$ to the environment $\hh_{E}$. In the
Schr\"odinger picture representation, it is obtained by tracing
out the output system $\hh_{B}$ instead of $\hh_{E}$:
\begin{equation}
    \label{eq:tradeoff02}
        T_{E*} (\varrho) = {\rm tr }_{B} \, V^{} \varrho V^{*}
        \qquad \Longleftrightarrow \qquad
        T_{*} (\varrho) = {\rm tr }_{E} \, V^{} \varrho V^{*}
\end{equation}
for all $\varrho \in \bhhstar{A}$. Henceforth, we will usually
write $T_{B}$ for the channel $T_{}$ to better distinguish it from
its complementary channel $T_{E}$.

The name {\em complementary channel} has been suggested by
I.~Devetak and P.~Shor in the course of their investigation of
quantum degradable channels \cite{DS03}. Recently A.~Holevo
\cite{Hol05} has shown that the classical channel capacity of a
quantum channel $T_{B}$ is additive iff the capacity of its
complementary channel $T_{E}$ is additive. Analogous results have
been obtained independently by C.~King {\em et al.} \cite{KMN+05}
(who chose the term {\em conjugate channels}).

Since two complementary channels share a common Stinespring
isometry, the continuity theorem relates the cb-norm distance
between two quantum channels to the cb-norm distance between the
complementary channels. The complementary channel of the noiseless
channel is completely depolarizing. The continuity theorem then
allows us to give a dimension-independent estimate for the
information-disturbance tradeoff in terms of quantum channels:
%%%%%%%%%%%%%%%%%%%%%%%%%%%%%%%%%%%%%%%%%%%%%%%%%%%%%%%%%%%%%%%%%%%%%%%%%%%%%%%%%%
\begin{theo}
    \label{theo:tradeoff}
    {\bf (Information-Disturbance Tradeoff)}\\
    Let $\hh_{A}$ and $\hh_{B}$ be finite-dimensional Hilbert
    spaces, and suppose that $T_{B} \mathpunct : \bhh{B} \rightarrow
    \bhh{A}$ is a quantum channel with Stinespring dilation
    $(\hh_{E}, V)$. Let $T_{E} \mathpunct : \bhh{E} \rightarrow
    \bhh{A}$ be the complementary channel, as defined in
    Eq.~(\ref{eq:tradeoff01}) above.
    We then have the following tradeoff estimate:
    \begin{equation}
        \label{eq:tradeoff03}
            \frac{1}{4} \inf_{D} \cb{T_{B} D_{} - \id_{A}}^2 \leq
            \cb{T_{E} - S_{}} \leq 2 \, \inf_{D} \cb{T_{B} D_{} -
            \id_{A}}^{\frac{1}{2}},
    \end{equation}
    where the infimum is over all decoding channels $D \mathpunct
    : \bhh{A} \rightarrow \bhh{B}$. In Eq.~(\ref{eq:tradeoff03}),
    $S \mathpunct : \bhh{E} \rightarrow \bhh{A}$ denotes a
    {\em completely depolarizing channel}, i.\,e.,
    \begin{equation}
        \label{eq:tradeoff04}
            S(e) := \tr{\sigma \, e} \, \1_{A} \quad \forall \; e \in
            \bhh{E}
    \end{equation}
    for some fixed quantum state $\sigma \in \bhhstar{E}$.
\end{theo}
%%%%%%%%%%%%%%%%%%%%%%%%%%%%%%%%%%%%%%%%%%%%%%%%%%%%%%%%%%%%%%%%%%%%%%%%%%%%%%%%%%

The interpretation of the tradeoff theorem is straightforward:
Whenever we may find a decoding channel $D$ such that almost all
the information can be retrieved from the output of the quantum
channel $T_{B}$, the norm difference $\cb{T_{B} D_{} - \id_{A}}$
will be small. By the right half of Eq.~(\ref{eq:tradeoff03}), we
may then conclude that the complementary channel $T_{E}$ is very
well approximated by a completely depolarizing channel $S$, and
thus releases almost no information to the environment.
Consequently, if a non-negligible amount of information escapes to
the environment, for instance by means of a measurement performed
by an eavesdropper, this will inevitably disturb the system.
Hence, in quantum physics there is ``no measurement without
perturbation". We know from Eq.~(\ref{eq:distance03}) that cb-norm
and operator norm are completely equivalent in the vicinity of the
noiseless channel. So any disturbance in the transmission can
always be detected locally.

On the other hand, if we are assured that the channel $T_{E}$ is
close to some depolarizing channel $S$ in cb-norm, the left half
of Eq.~(\ref{eq:tradeoff03}) guarantees that we may find a
decoding channel $D$ which retrieves almost all the information
from the $B$-branch of the system. Consequently, there is ``no
perturbation without measurement". However, in this case it is
usually not enough to verify that $T_{E}$ erases information
locally; the channel also needs to destroy correlations. We will
come back to this distinction and its implications for the
interpretation of
the tradeoff theorem in Sec.~\ref{sec:alternative}.\\
\\
%%%%%%%%%%%%%%%%%%%%%%%%%%%%%%%%%%%%%%%%%%%%%%%%%%%%%%%%%%%%%%%%%%%%%%%%%%%%%%%%%%
{\bf Proof of Th.~\ref{theo:tradeoff}:} It is easily verified that
a Stinespring isometry for the completely depolarizing channel $S
\mathpunct : \bhh{E} \rightarrow \bhh{A}$, as given in
Eq.~(\ref{eq:tradeoff04}), is the isometric embedding
\begin{equation}
    \label{eq:tradeoff05}
            V_{S} \mathpunct : \hh_{A}  \rightarrow \hh_{A}
            \otimes \hh_{E'} \otimes \hh_{E} \qquad
            \ket{\varphi} \mapsto \ket{\varphi} \otimes
            \ket{\psi_{\sigma}},
\end{equation}
where $\hh_{E'} \cong \hh_{E}$, and $\ket{\psi_{\sigma}} \in
\hh_{E'} \otimes \hh_{E}$ is a purification of $\sigma \in
\bhhstar{E}$. Thus, the completely depolarizing channel $S_{}
\equiv S_{E'E}$ and the ideal channel $\id_{A}$ are indeed
complementary.

The tradeoff theorem is then a straightforward consequence of the
continuity theorem. Let us focus on the left half of
Eq.~(\ref{eq:tradeoff03}) first, and assume that $V_{T} \mathpunct
: \hh_{A} \rightarrow \hh_{B} \otimes \hh_{E}$ is a Stinespring
dilation for the quantum channel $T_{E}$ (and its complementary
channel $T_{B}$, respectively). Let $V_{S} \mathpunct : \hh_{A}
\rightarrow \hh_{A} \otimes \hh_{E'} \otimes \hh_{E}$ be the
Stinespring isometry of $S_{E'E}$ given by
Eq.~(\ref{eq:tradeoff05}). Note that the dilation spaces $\hh_{B}$
and $\hh_{A} \otimes \hh_{E'}$ are not necessarily of the same
size. However, we can easily correct for this by suitably
enlarging the smaller system, $\hh_{B}$ say. The left half of the
continuity estimate Eq.~(\ref{eq:continuity04}) then guarantees
the existence of an isometry $V_{D} \mathpunct : \hh_{B}
\rightarrow \hh_{A} \otimes \hh_{E'}$ such that
\begin{equation}
    \label{eq:tradeoff06}
        \norm{ (V_{D} \otimes \1_{E}) V_{T} - V_{S}} \leq
        \cb{T_{E} - S_{E'E}}^{\frac{1}{2}}.
\end{equation}
As illustrated in Fig.~\ref{fig:tradeoff01}, the isometry $V_{D}$
defines a decoding channel
\begin{equation}
    \label{eq:tradeoff07}
        D_{A} \mathpunct : \bhh{A} \rightarrow \bhh{B} \qquad
        D_{A}(a) := V_{D}^{*} (a \otimes \1_{E'}) V_{D}^{},
\end{equation}
and by the right half of the continuity estimate
Eq.~(\ref{eq:continuity04}) we may now conclude that
\begin{equation}
    \label{eq:tradeoff08}
        \cb{T_{B} D_{A} - \id_{A}} \leq 2 \,
        \cb{T_{E} - S_{E'E}}^{\frac{1}{2}},
\end{equation}
which proves the left half of Eq.~(\ref{eq:tradeoff03}).
$\blacktriangle$
%%%%%%%%%%%%%%%%%%%%%%%%%%%%%%%%%%%%%%%%%%%%%%%%%%%%%%%%%%%%%%%%%%%%%%%%%%%%%%%%%%
\begin{figure}
    \begin{center}
        \psfrag{E}[cc][cc]{$\hh_{E}$}
        \psfrag{A}[cc][cc]{$\hh_{A}$}
        \psfrag{B}[cc][cc]{$\hh_{B}$}
        \psfrag{F}[cc][cc]{$\hh_{E'}$}
        \psfrag{T}[bc][Bc]{{\Large $V_{T}$}}
        \psfrag{D}[cc][cr]{{\Large $V_{D}$}}
        \psfrag{p}[Bc][cc]{$\ket{\varphi}$}
        \psfrag{q}[Bc][cc]{$\approx \ket{\varphi}$}
        \psfrag{s}[cc][cc]{$\approx \ket{\psi_{\sigma}}$}
        \includegraphics*[width=0.7\columnwidth]{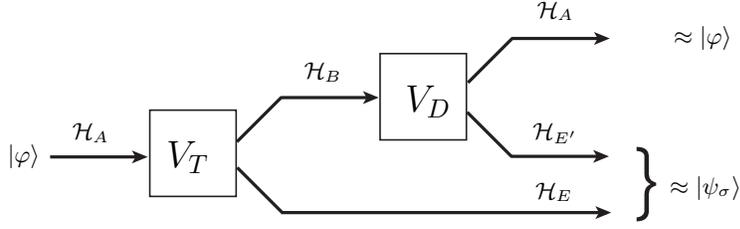}
        \caption{\label{fig:tradeoff01} {\it ``No
        perturbation without measurement." Whenever the cb-norm difference
        $\cb{T_{E} - S_{E'E}}$ is small, we may find a decoding channel
        $D_{A}$ with Stinespring isometry $V_{D}$ such that
        the concatenated isometry $(V_{D}
        \otimes \1_{E}) V_{T}$ hardly differs from the Stinespring
        isometry of a noiseless channel, with some fixed
        $\ket{\psi_{\sigma}} \in \hh_{E'} \otimes \hh_{E}.$}}
    \end{center}
\end{figure}
%%%%%%%%%%%%%%%%%%%%%%%%%%%%%%%%%%%%%%%%%%%%%%%%%%%%%%%%%%%%%%%%%%%%%%%%%%%%%%%%%%

The proof of the right half of Eq.~(\ref{eq:tradeoff03}) proceeds
very much along the same lines: Assume that $V_{T} \mathpunct :
\hh_{A} \rightarrow \hh_{B} \otimes \hh_{E}$ and $V_{D} \mathpunct
: \hh_{B} \rightarrow \hh_{A} \otimes \hh_{E'}$ are Stinespring
isometries for the quantum channels $T_{B}$ and $D_{A}$,
respectively. As in Eq.~(\ref{eq:tradeoff05}), let $V_{S}
\mathpunct : \hh_{A} \rightarrow \hh_{A} \otimes \hh_{E'} \otimes
\hh_{E}$ denote the Stinespring isometry of the ideal channel
$\id_{A}$ and its complementary channel, the completely
depolarizing channel $S_{E'E}$. Just as before, the left half of
the continuity estimate Eq.~(\ref{eq:continuity04}) assures us
that we may find a unitary operator $U \in \bhh{E'E}$ such that
(cf. Fig.~\ref{fig:tradeoff02})
\begin{equation}
    \label{eq:tradeoff09}
        \norm{(\1_{A} \otimes U_{}) (V_{D} \otimes \1_{E}) V_{T} -
        V_{S}} \leq \cb{T_{B} D_{A} - \id_{A}}^{\frac{1}{2}},
\end{equation}
where again we have suitably enlarged the dilation space
$\hh_{E'}$, if necessary.
%%%%%%%%%%%%%%%%%%%%%%%%%%%%%%%%%%%%%%%%%%%%%%%%%%%%%%%%%%%%%%%%%%%%%%%%%%%%%%%%%%
\begin{figure}
    \begin{center}
        \psfrag{E}[cc][cc]{$\hh_{E}$}
        \psfrag{A}[cc][cc]{$\hh_{A}$}
        \psfrag{B}[cc][cc]{$\hh_{B}$}
        \psfrag{F}[cc][cc]{$\hh_{E'}$}
        \psfrag{T}[bc][Bc]{{\Large $V_{T}$}}
        \psfrag{D}[cc][cr]{{\Large $V_{D}$}}
        \psfrag{U}[cc][cc]{{\Large $U$}}
        \psfrag{p}[Bc][cc]{$\ket{\varphi}$}
        \psfrag{q}[Bc][cc]{$\approx \ket{\varphi}$}
        \psfrag{s}[cc][cc]{$\approx \ket{\psi_{\sigma}}$}
        \includegraphics*[width=0.9\columnwidth]{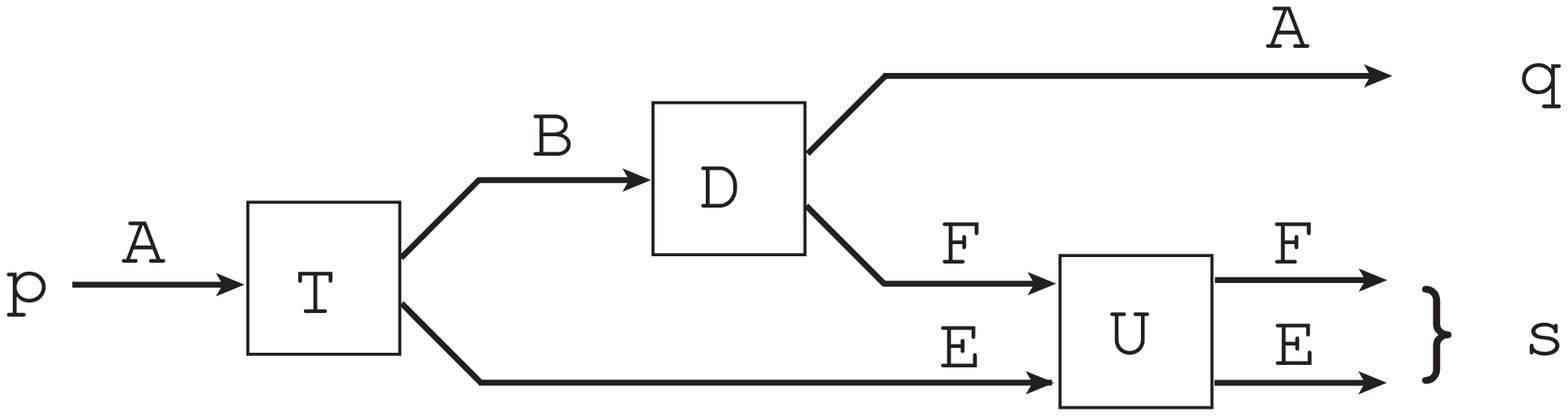}
        \caption{\label{fig:tradeoff02} {\it ``No
        measurement without perturbation." If there exists a decoding
        channel $D_{A}$ with Stinespring isometry $V_{D}$ such that
        the cb-norm difference
        $\cb{T_{B} D_{A} - \id_{A}}$ is small, we may find a
        unitary $U \in \bhh{E'} \otimes \bhh{E}$ such that the
        concatenated isometry $(\1_{A} \otimes U) (V_{D} \otimes
        \1_{E}) V_{T}$ is close to the Stinespring
        isometry of a completely depolarizing channel, with some fixed
        $\ket{\psi_{\sigma}} \in \hh_{E'} \otimes \hh_{E}.$}}
    \end{center}
\end{figure}
%%%%%%%%%%%%%%%%%%%%%%%%%%%%%%%%%%%%%%%%%%%%%%%%%%%%%%%%%%%%%%%%%%%%%%%%%%%%%%%%%%

Setting ${\rm Ad}_{V_{T}} := V_{T}^{*} (\cdot) V_{T}^{}$ and ${\rm
Ad}_{U^{*}} := U^{} (\cdot) U^{*}$, we may now conclude from the
right half of Eq.~(\ref{eq:continuity04}) that
\begin{equation}
    \label{eq:tradeoff10}
        \cb{{\rm Ad}_{V_{T}} \circ (D_{E'} \otimes \id_{E}) - S_{E'E} \circ {\rm
        Ad}_{U^{*}}} \leq 2 \, \cb{T_{B} D_{A} -
        \id_{A}}^{\frac{1}{2}},
\end{equation}
which is almost the desired result. It only remains to restrict
the depolarizing channel $S_{E'E}$ to the $E$-branch of the output
system. Obviously, since $S_{E'E}$ is completely depolarizing on
the combined output system $\hh_{E'} \otimes \hh_{E}$, the same
holds true after a unitary rotation by $U^{*}$ and the restriction
to one of the branches. In particular, by setting
\begin{equation}
    \label{eq:tradeoff11}
        \widetilde{S}_{E} \mathpunct : \bhh{E} \rightarrow \bhh{A}
        \qquad \widetilde{S}_{E} (e) := S_{E'E} \circ {\rm Ad}_{U^{*}}
        (\1_{E'} \otimes e)
\end{equation}
we obtain a completely depolarizing channel on the restricted
system $\hh_{E}$ such that
\begin{equation}
    \label{eq:tradeoff12}
        \widetilde{S}_{E} (e) = \tr{\widetilde{\sigma} \, e} \, \1_{A}
        \quad \forall \; \; e \in \bhh{E}
\end{equation}
for $\widetilde{\sigma} := {\rm tr}_{E'} \, U^{*}
\kb{\psi_{\sigma}} U^{}$. It then immediately follows from
Eq.~(\ref{eq:tradeoff10}) that
\begin{equation}
    \label{eq:tradeoff13}
        \cb{T_{E} - \widetilde{S}_{E}} \leq 2 \, \cb{T_{B} D_{A} -
        \id_{A}}^{\frac{1}{2}},
\end{equation}
as advertised. $\blacksquare$\\
%%%%%%%%%%%%%%%%%%%%%%%%%%%%%%%%%%%%%%%%%%%%%%%%%%%%%%%%%%%%%%%%%%%%%%%%%%%%%%%%%%

The tradeoff theorem amounts to a simple continuity estimate for
the no-broadcasting and no-cloning theorems: A quantum channel $T
\mathpunct : \bhh{1} \otimes \bhh{2} \rightarrow \bh$ with a
triple of isomorphic Hilbert spaces $\hh_{1} \cong \hh_{2} \cong
\hh$ is said to {\em broadcast} the quantum state $\varrho \in
\bhstar$ iff the restrictions of the output state $T_{*}
(\varrho)$ to both subsystems coincide with the input $\varrho$,
${\rm tr_{2}} \, T_{*} (\varrho) = \varrho = {\rm tr_{1}} \, T_{*}
(\varrho).$ The only way to broadcast a pure state $\varrho =
\kb{\psi}$ is to generate the product state $\kb{\psi} \otimes
\kb{\psi}$. Thus, broadcasting of pure states is equivalent to
cloning. H.~Barnum {\it et al.} \cite{BCF+96} have shown that a
quantum channel $T$ can broadcast two quantum states $\varrho_{1}$
and $\varrho_{2}$ iff they commute --- an extension of the famous
no-cloning theorem \cite{WZ82,Die82} to mixed states.

The tradeoff theorem immediately shows that approximate
broadcasting is also impossible, and provides
dimension-independent bounds:
%%%%%%%%%%%%%%%%%%%%%%%%%%%%%%%%%%%%%%%%%%%%%%%%%%%%%%%%%%%%%%%%%%%%%%%%%%%%%%%%%%
\begin{coro}
    \label{coro:broadcast}
        {\bf (No Broadcasting)}\\
        Let $V \mathpunct : \hh_{} \rightarrow \hh_{1} \otimes \hh_{2}
        \otimes \hh_{E}$ be a Stinespring isometry for the quantum
        channel $T \mathpunct : \bhh{1} \otimes \bhh{2} \rightarrow \bh$ with local
        restrictions $T_{1} (a) := V^{*} (a \otimes
        \1_{2} \otimes \1_{E}) V^{}$ and $T_{2} (a) := V^{*} (\1_{1}
        \otimes a \otimes \1_{E}) V^{}$. Then there exists a
        completely depolarizing channel $S
        \mathpunct : \bhh{2} \rightarrow \bh$ defined as in
        Eq.~(\ref{eq:tradeoff04}) such that
        \begin{equation}
           \label{eq:tradeoff15}
                \cb{T_{2} - S_{}} \leq 2 \, \cb{T_{1} -
                \id_{}}^{\frac{1}{2}}.
        \end{equation}
\end{coro}
%%%%%%%%%%%%%%%%%%%%%%%%%%%%%%%%%%%%%%%%%%%%%%%%%%%%%%%%%%%%%%%%%%%%%%%%%%%%%%%%%%

Hence, any broadcast channel that has reasonably high fidelity in
one of the output branches releases little information to the
other branch (and the environment). While
Corollary~\ref{coro:broadcast} shows that neither perfect nor
approximate broadcasting is possible, the bound is certainly not
tight. The merit of the tradeoff theorem is a
dimension-independent estimate, while optimal cloning bounds are
known to depend strongly on the dimension of the underlying
Hilbert space \cite{SIG05,CF05}.

%%%%%%%%%%%%%%%%%%%%%%%%%%%%%%%%%%%%%%%%%%%%%%%%%%%%%%%%%%%%%%%%%%%%%%%%%%%%%%%%%%
%%%%%%%%%%%%%%%%%%%%%%%%%%%%%%%%%%%%%%%%%%%%%%%%%%%%%%%%%%%%%%%%%%%%%%%%%%%%%%%%%%
%%%%%%%%%%%%%%%%%%%%%%%%%%%%%%%%%%%%%%%%%%%%%%%%%%%%%%%%%%%%%%%%%%%%%%%%%%%%%%%%%%
%%%%%%%%%%%%%%%%%%%%%%%%%%%%%%%%%%%%%%%%%%%%%%%%%%%%%%%%%%%%%%%%%%%%%%%%%%%%%%%%%%

\section{Weaker Notions of Disturbance and Erasure}
    \label{sec:alternative}

The tradeoff estimate established in Sec.~\ref{sec:tradeoff} has
the somewhat surprising and very welcome feature of being
completely independent of the dimensions of the underlying Hilbert
spaces, which makes it ideally suited for applications in which
these dimensions are unknown and possibly very large, as in black
hole evaporation. However, this property depends crucially on the
choice of the distance measure: in this Section we will give an
example of a quantum channel $T \mathpunct : \bc{\nu} \rightarrow
\bc{\nu}$ such that $\norm{T - S} \approx 0$ for $\nu \to \infty$,
but $\cb{T - S} \geq 1.$ The example shows that operator norm and
cb-norm are in general inequivalent in the vicinity of the
completely depolarizing channel $S$. In contrast, we know from
Eq.~(\ref{eq:distance03}) that equivalence does hold in the
neighborhood of the noiseless channel.

An example for a channel which nicely demonstrates this separation
is
\begin{equation}
    \label{eq:alternative02}
        T \mathpunct : \bc{\nu} \rightarrow \bc{\nu} \qquad T:=
        \frac{\nu}{\nu+1} S + \frac{1}{\nu+1} \Theta,
\end{equation}
where $S \mathpunct : \bc{\nu} \rightarrow \bc{\nu}$ is the
completely depolarizing channel given again by
\begin{equation}
    \label{eq:alternative03}
        S (e) = \frac{1}{\nu} \tr{e} \, \1 \quad \forall \; e \in \bc{\nu}
        \qquad \Longleftrightarrow \qquad S_{*} (\varrho) = \frac{1}{\nu}
        \1 \quad \forall \; \varrho \in \bcstar{\nu},
\end{equation}
and $\Theta \mathpunct : \bc{\nu} \rightarrow \bc{\nu}$ is the
so-called {\em transpose map:} $\Theta (e) = e^{t}$, the matrix
transpose of $e \in \bc{\nu}$. While $\Theta$ is linear, unital
and positive, it is not completely positive, and thus cannot be
implemented as a quantum channel \cite{Pau02}. However, we will
show in Prop.~\ref{propo:transpose} below that $T$ nonetheless
remains a valid quantum channel.

Noting that $\norm{\Theta} = 1$ and $\cb{\Theta} = \nu$
\cite{Pau02}, it then immediately follows that
\begin{equation}
    \label{eq:alternative04}
        \norm{T - S} = \frac{1}{\nu + 1} \, \norm{\Theta - S} \leq
        \frac{1}{\nu + 1} \, \big ( \norm{\Theta} + \norm{S} \big ) =
        \frac{2}{\nu + 1},
\end{equation}
and thus $\lim_{\nu\to\infty} \norm{T-S} = 0$. On the other hand,
making again use of the triangle inequality we have the lower
bound
\begin{equation}
    \label{eq:alternative05}
        \cb{T - S} = \frac{1}{\nu + 1} \, \cb{\Theta - S} \geq
        \frac{1}{\nu + 1} \, \big ( \cb{\Theta} - \cb{S} \big ) =
        \frac{\nu - 1}{\nu + 1}.
\end{equation}
This demonstrates the suggested separation between cb-norm and
operator norm as $\nu \to \infty$. It only remains to show that
$T$ is a quantum channel, which will become clear from the proof
of
%%%%%%%%%%%%%%%%%%%%%%%%%%%%%%%%%%%%%%%%%%%%%%%%%%%%%%%%%%%%%%%%%%%%%%%%%%%%%%%%%%
\begin{propo}
    \label{propo:transpose}
        Let $\Theta \mathpunct : \bc{\nu} \rightarrow \bc{\nu}$ be
        the transpose map, and let $S \mathpunct : \bc{\nu} \rightarrow
        \bc{\nu}$ be the completely depolarizing channel as given
        in Eq.~(\ref{eq:alternative03}). Then
        \begin{equation}
            \label{eq:alternative06}
                T_{p} \mathpunct : \bc{\nu} \rightarrow \bc{\nu}
                \qquad T_{p} := (1-p) \, S + p \, \Theta
        \end{equation}
        for $p \in [0,1]$ defines a quantum channel iff $p \leq
        \frac{1}{\nu + 1}$.
\end{propo}
%%%%%%%%%%%%%%%%%%%%%%%%%%%%%%%%%%%%%%%%%%%%%%%%%%%%%%%%%%%%%%%%%%%%%%%%%%%%%%%%%%
{\bf Proof:} While $T_{p}$ is clearly linear, unital and positive
for all $p \in [0,1]$, it is not necessarily completely positive,
since the transpose map $\Theta$ does not have this property. In
order to test for complete positivity, it is sufficient to apply
the Schr\"odinger dual $T_{p*}$ to half of a maximally entangled
state $\ket{\Omega} := \frac{1}{\sqrt{\nu}} \sum_{i=1}^{\nu}
\ket{i, i}$ on $\C^{\nu} \otimes \C^{\nu}$. In fact, it follows
from Jamiolkowski's duality theorem (cf. \cite{Jam72} and
Th.~2.3.4 in \cite{Key02}) that a linear map $R \mathpunct :
\bc{\nu} \rightarrow \bc{\nu}$ is completely positive iff $\varrho
:= R_{*} \otimes \id \kb{\Omega}$ is a quantum state.

We will now apply this statement to the family $T_{p}$. It is
easily seen from Eq.~(\ref{eq:dual}) that the Schr\"odinger dual
$\Theta_{*}$ coincides with $\Theta_{}$, $\Theta_{*} = \Theta_{}$.
Straightforward calculation shows that
\begin{equation}
    \label{eq:alternative08}
        \varrho_{p} := T_{p*} \otimes \id \kb{\Omega} =
        \frac{1-p}{\nu^2} \, \1 \otimes \1 + \frac{p}{\nu} \, \F,
\end{equation}
where $\F := \sum_{i,j} \op{i,j}{j,i}$ is the so-called {\em flip
operator}. Note that $\F^{*} = \F$ and $\F^2 = \1$. Quantum states
of the form
\begin{equation}
    \label{eq:alternative09}
        \varrho = \alpha \, \1 + \beta \, \F
\end{equation}
are usually called {\em Werner states} \cite{Wer89}. In order to
see for which values of the parameters $\alpha$ and $\beta$ the
operator $\varrho$ describes a quantum state, it is useful to
rewrite Eq.~(\ref{eq:alternative09}) in terms of the
eigenprojections $P_{\pm}$ of the flip operator $\F$, i.\,e.,
$\F_{} \, P_{\pm} \, \ket{\psi} = \pm \, P_{\pm} \ket{\psi}$ with
\begin{equation}
    \label{eq:alternative10}
        P_{+} := \frac{\1 + \F}{2} \qquad {\rm and} \qquad P_{-}
        := \frac{\1 - \F}{2}.
\end{equation}
$P_{+}$ is the projection onto the symmetric (Bose) subspace,
while $P_{-}$ describes the projection onto the antisymmetric
(Fermi) subspace. Observing that $P_{+} + P_{-} = \1_{}$ and
$P_{+} - P_{-} = \F_{}$ and substituting these expressions into
Eq.~(\ref{eq:alternative09}), we see that
\begin{equation}
    \label{eq:alternative11}
        \varrho = (\alpha + \beta) \, P_{+} + (\alpha - \beta) \,
        P_{-},
\end{equation}
which is positive iff $\alpha \geq \beta$. This implies that the
output state $\varrho_{p}$, as given in
Eq.~(\ref{eq:alternative08}), is a quantum state (and thus $T_{p}$
is completely positive, by the Jamiolkowski duality) iff $p
\leq \frac{1}{\nu+1}$, as suggested. $\blacksquare$\\
%%%%%%%%%%%%%%%%%%%%%%%%%%%%%%%%%%%%%%%%%%%%%%%%%%%%%%%%%%%%%%%%%%%%%%%%%%%%%%%%%%

Hayden {\it et al.} \cite{HLS+04} have recently proven that random
selections of unitary matrices generically show an even stronger
separation: in their terminology, a quantum channel $R \mathpunct
: \bc{\nu} \rightarrow \bc{\nu}$ is called {\em
$\varepsilon$-randomizing} iff
\begin{equation}
    \label{eq:alternative13}
        \norm{R_{*}(\varrho) - \frac{1}{\nu} \, \1} \leq
        \frac{\varepsilon}{\nu} \quad \forall \; \; \varrho \in
        \bcstar{\nu}.
\end{equation}
For sufficiently large $\nu$, Hayden {\it et al.} show that a
random selection of $\mu \sim \frac{1}{\varepsilon^2} \nu \log
\nu$ unitary operators $\{ U_{i} \}_{i=1}^{\mu} \subset \bc{\nu}$
with high probability yields an $\epsilon$-randomizing quantum
channel,
\begin{equation}
    \label{eq:alternative14}
        R(e) := \frac{1}{\mu} \, \sum_{i=1}^{\mu} \,
        U_{i}^{*} \, e \, U_{i}^{} \quad \forall \; \;
        e \in \bc{\nu}.
\end{equation}
In striking contrast, exact randomization of quantum states (such
that $\varepsilon = 0$ in Eq.~(\ref{eq:alternative13})) is known
to require an ancilla system of dimension at least $\nu^2 \gg \mu$
\cite{AMT+00}.

The definition of approximate randomization given in
Eq.~(\ref{eq:alternative13}) implies the strictly weaker estimate
$\norm{R - S} \leq \varepsilon$, with the completely depolarizing
channel $S$ as in Eq.~(\ref{eq:alternative03}) above. However,
Eq.~(\ref{eq:alternative13}) does not amount to the stabilized
version $\cb{R - S} \leq \varepsilon$. Hayden {\it et al.}
explicitly demonstrate this by showing that for
$\varepsilon$-randomizing channels $R$ randomly generated as in
Eq.~(\ref{eq:alternative14}) above, one always has the upper bound
\begin{equation}
    \label{eq:alternative15}
        \tracenorm{(R_{*} - S_{*}) \otimes \id (\kb{\Omega})} \geq 2 \, \big (
        1 - \frac{\mu}{\nu^2} \big ) \xrightarrow{\nu \to \infty}
        2,
\end{equation}
where $\ket{\Omega} := \frac{1}{\sqrt{\nu}} \sum_{i=1}^{\nu}
\ket{i, i}$ again denotes the maximally entangled state on
$\C^{\nu} \otimes \C^{\nu}$.

The bound in Eq.~(\ref{eq:alternative15}) implies that
$\lim_{\nu\to\infty} \cb{R-S} = 2$, and the same holds true for
any other channel $R$ with an ancilla system of dimension
$o(\nu^2)$. From the right half of the tradeoff theorem
Eq.~(\ref{eq:tradeoff03}) we may then conclude that for none of
these channels will it be possible to find a decoding channel $D$
such that the randomized information can be recovered from the
ancilla system alone. Information may remain hidden in quantum
correlations and cannot be retrieved locally.\\

Note that while these examples demonstrate that it is in general
not possible to upper bound the cb-norm $\cb{\cdot}$ in terms of
the operator norm $\norm{\cdot}$ with a dimension-independent
estimate, a dimension-{\em dependent} bound can of course be
given. In fact, for any linear map $R \mathpunct : \A \rightarrow
\bc{\nu}$ with an arbitrary (possibly infinite) $C^{*}$-algebra
$\A$ we have $\cb{R} \leq \nu \, \norm{R}$ \cite{Pau02}. The
transpose map $\Theta$ shows that this bound can be tight.

%%%%%%%%%%%%%%%%%%%%%%%%%%%%%%%%%%%%%%%%%%%%%%%%%%%%%%%%%%%%%%%%%%%%%%%%%%%%%%%%%%
%%%%%%%%%%%%%%%%%%%%%%%%%%%%%%%%%%%%%%%%%%%%%%%%%%%%%%%%%%%%%%%%%%%%%%%%%%%%%%%%%%
%%%%%%%%%%%%%%%%%%%%%%%%%%%%%%%%%%%%%%%%%%%%%%%%%%%%%%%%%%%%%%%%%%%%%%%%%%%%%%%%%%
%%%%%%%%%%%%%%%%%%%%%%%%%%%%%%%%%%%%%%%%%%%%%%%%%%%%%%%%%%%%%%%%%%%%%%%%%%%%%%%%%%

\section{Further Applications}
    \label{sec:further}

We have shown in Sec.~\ref{sec:tradeoff} how the continuity
theorem entails dimension-independent bounds for the
information-disturbance tradeoff in terms of stabilized operator
norms. These results go beyond the standard measurement-based
approach and can be seen as complementary to the entropic bounds
obtained recently by Christandl and Winter \cite{CW05}: Assume
that a uniform quantum ensemble $E_{1} := \{ \frac{1}{\nu},
\ket{i} \}$ of basis states of the Hilbert space $\hh \cong
\C^{\nu}$ and the Fourier-rotated ensemble $E_{2} := \{
\frac{1}{\nu}, U \ket{i} \}$ have both nearly maximal Holevo
information when sent through the quantum channel $T_{B}
\mathpunct : \bc{\nu} \rightarrow \bc{\nu}$:
\begin{equation}
    \label{eq:further01}
        \chi \big ( T_{B} (E_{k}) \big ) \geq \ld \nu - \varepsilon
\end{equation}
for $k = 1, 2$ and some (small) $\varepsilon > 0$, where $\ld \nu$
is the dual logarithm of $\nu$ and
\begin{equation}
    \label{eq:further02}
        \chi \big ( T_{B} (E_{1}) \big ) := S \bigg ( \frac{1}{\nu}
        \sum_{i=1}^{\nu} T_{B*} \big ( \kb{i} \big ) \bigg ) -
        \frac{1}{\nu} \sum_{i=1}^{\nu} S
        \Big ( T_{B*} \big ( \kb{i} \big ) \Big )
\end{equation}
denotes the Holevo information of the output ensemble $T_{B}
(E_{1}) := \{ \frac{1}{\nu}, T_{B*} (\kb{i}) \}$. Analogously,
$\chi \big ( T_{B} (E_{2}) \big)$ is the Holevo information of the
rotated ensemble $T_{B} (E_{2}) := \{ \frac{1}{\nu}, T_{B*}
(U^{}\kb{i}U^{*}) \}$. Christandl and Winter then conclude from
their entropic uncertainty relation that the {\em coherent
information}
\begin{equation}
    \label{eq:further03}
        I_{c} \Big ( T_{B}, \frac{1}{\nu} \1 \Big ) := S \bigg (
        \frac{1}{\nu} \sum_{i=1}^{\nu} T_{B} \big ( \kb{i} \big )
        \bigg) - S \Big ( T_{B*} \otimes \id \big ( \kb{\Omega}
        \big ) \Big ) \geq \ld \nu - 2 \varepsilon
\end{equation}
is likewise large, where again $\ket{\Omega} :=
\frac{1}{\sqrt{\nu}} \sum_{i=1}^{\nu} \ket{i, i}$ denotes a
maximally entangled state on $\C^{\nu} \otimes \C^{\nu}$. As a
consequence, there exists a decoding operation $D \mathpunct :
\bc{\nu} \rightarrow \bc{\nu}$ such that $F_{c} (T_{B}D_{}) \geq 1
- 2 \sqrt{2 \varepsilon}$ \cite{SW02}, where $F_{c}(R)$ denotes
the {\em channel fidelity} of the quantum channel $R$, $F_{c} (R)
:= \bra{\Omega} R_{*} \otimes \id ( \kb{\Omega} ) \ket{\Omega}$.
However, the faithful transmission of the maximally entangled
state $\ket{\Omega}$ alone is not sufficient to conclude that
$T_{B} D_{} \approx \id$ in operator norm with
dimension-independent bounds \cite{KW04}. But it is always
possible to find a subspace $\hh' \subset \hh$ with $\dim \hh'
\geq \frac{1}{2} \dim \hh$ such that $\cb{T_{B}^{'}D_{}^{'}-\id'}
\leq 13 \, \varepsilon^{\frac{1}{8}}$ \cite{BKN00,KW04}, where
$T_{B}^{'}$ and $D_{}^{'}$ are channels whose range and domain,
respectively, are restricted to $\hh'$. The tradeoff theorem then
guarantees that $\cb{T_{E}^{'} - S} \leq 8 \,
\varepsilon^{\frac{1}{16}}$ for some completely depolarizing
channel $S$. Thus, in combination with Th.~\ref{theo:tradeoff} the
existence of highly reliable detectors for a basis and its
conjugate alone imply a stabilized version of privacy, which is in
general much stronger than the entropic version that appears in
\cite{CW05}. The improvement comes at the expense of a smaller
code space. However, in many applications this is an exponentially
large space, hence its reduction by a factor $1/2$ does not affect
the rate of the protocol.

The information-disturbance tradeoff also plays the central role
in the infamous black hole information loss puzzle: black holes
emit thermal Hawking radiation \cite{Haw74}, which contains
(almost) no information about the previously absorbed quantum
states. Hawking's derivation is perturbative, but we can
nonetheless try to model this evaporation process as an (almost)
completely depolarizing quantum channel, $T_{E} \approx S_{}$. The
tradeoff theorem then suggests that all the data about the
formation of the black hole reside inside the event horizon, and
could at least in principle be retrieved from there. However, the
black hole may eventually evaporate completely, seemingly erasing
all this information in the process and hence violating the
unitarity of quantum mechanics.

The tradeoff theorem provides the explicit bounds for this
estimate. Our results also show that for large systems with many
internal degrees of freedom --- such as all the information
swallowed by a black hole ---, the estimate crucially depends on
the choice of the operator topology. If only an unstabilized
estimate $\norm{T_{E} - S_{}} \leq \varepsilon$ can be guaranteed,
information may remain hidden in quantum correlations between the
thermal radiation and the black hole final state.

Similar conclusions apply to thermalization processes, in which a
quantum system approaches an equilibrium state via repeated
interaction with an environment. In so-called {\em collision
models} \cite{SZS+02,ZSB+02} the evolution of the thermalizing
quantum system is described in terms of a quantum channel $T_{E}$.
If $T_{E}$ is almost completely depolarizing in cb-norm, all the
information about the initial state of the system will have
dissipated into the environment, and can at least in principle be
retrieved from there.

Braunstein and Pati \cite{BP05} have explored the consequences of
the information-disturbance tradeoff for the physics of black
holes and thermalization in greater detail.

%%%%%%%%%%%%%%%%%%%%%%%%%%%%%%%%%%%%%%%%%%%%%%%%%%%%%%%%%%%%%%%%%%%%%%%%%%%%%%%%%%
%%%%%%%%%%%%%%%%%%%%%%%%%%%%%%%%%%%%%%%%%%%%%%%%%%%%%%%%%%%%%%%%%%%%%%%%%%%%%%%%%%
%%%%%%%%%%%%%%%%%%%%%%%%%%%%%%%%%%%%%%%%%%%%%%%%%%%%%%%%%%%%%%%%%%%%%%%%%%%%%%%%%%
%%%%%%%%%%%%%%%%%%%%%%%%%%%%%%%%%%%%%%%%%%%%%%%%%%%%%%%%%%%%%%%%%%%%%%%%%%%%%%%%%%

\section{Conclusions}
    \label{sec:summary}

In conclusion, we have shown and explored a continuity theorem for
Stinespring's dilation theorem: two quantum channels, $T_{1},
T_{2}$ are close in cb-norm iff there exist corresponding
Stinespring isometries, $V_{1}, V_{2}$, which are close in
operator norm.

When applied to the noiseless channel $T_{1} = \id_{}$, the
continuity theorem yields a formulation of the
information-disturbance tradeoff in which both information gain
and disturbance are measured in terms of operator norms,
complementing recently obtained entropic bounds.

In the form we have presented it, the continuity theorem applies
to quantum channels on finite-dimensional quantum systems and
yields dimension-independent bounds. This makes the result ideally
suited for applications to situations in which these dimensions
are large or possibly unknown.

The absence of dimension-dependent factors in the continuity
bounds Eq.~(\ref{eq:continuity04}) seems to indicate that the
result is not restricted to the finite-dimensional setting.
Extensions of the continuity theorem to completely positive maps
between arbitrary $C^{*}$-algebras are currently under
investigation.

Infinite dimensions lead to a number of complications. The neat
and useful one-to-one correspondence between states and density
operators fails to hold in infinite-dimensional Hilbert spaces:
there are always positive linear functionals on $\bh$ which cannot
be represented as trace class operators \cite{Seg47}. The dual
space of $\bhstar$ is the space of compact operators on $\bh$, not
the full operator space $\bh$. The states that do allow a tracial
representation are called {\em normal}. A positive functional
$\omega \mathpunct : \bh \rightarrow \C$ is normal iff
$\lim_{n\to\infty} \omega(b_n) = \omega (b)$ for every sequence
$(b_{n})_{n}$ of norm-bounded increasing operators with least
upper bound $b \in \bh$ (cf. \cite{Dav76}, Ch.~1.6). More
generally, a quantum channel $T \mathpunct : \bhh{B} \rightarrow
\bhh{A}$ is {\em normal} iff $\lim_{n\to\infty} T(b_{n}) = T(b)$.
The normal channels $T$ are then precisely those for which the
duality relation Eq.~(\ref{eq:dual}) continues to hold (cf.
\cite{Dav76}, Ch.~9). Non-normal (or {\em singular}) channels do
not have a Schr\"odinger dual.

However, as long as the Hilbert spaces are separable and all
systems obey generic energy constraints, the state space will be
compact \cite{Hol03}, the channels respecting these energy
constraints will be normal, and our proof of the continuity
theorem and the tradeoff bounds then goes through unchanged. Thus,
all the results presented in this work continue to hold in the
practically relevant settings.

%%%%%%%%%%%%%%%%%%%%%%%%%%%%%%%%%%%%%%%%%%%%%%%%%%%%%%%%%%%%%%%%%%%%%%%%%%%%%%%%%%
%%%%%%%%%%%%%%%%%%%%%%%%%%%%%%%%%%%%%%%%%%%%%%%%%%%%%%%%%%%%%%%%%%%%%%%%%%%%%%%%%%
%%%%%%%%%%%%%%%%%%%%%%%%%%%%%%%%%%%%%%%%%%%%%%%%%%%%%%%%%%%%%%%%%%%%%%%%%%%%%%%%%%
%%%%%%%%%%%%%%%%%%%%%%%%%%%%%%%%%%%%%%%%%%%%%%%%%%%%%%%%%%%%%%%%%%%%%%%%%%%%%%%%%%

\begin{acknowledgments}
We would like to thank S. Braunstein, M. Christandl, A. Holevo,
and A. Winter for fruitful and stimulating discussions.

DK is grateful for financial support from the European Union
project RESQ and the German Academic Exchange Service (DAAD).
\end{acknowledgments}

%%%%%%%%%%%%%%%%%%%%%%%%%%%%%%%%%%%%%%%%%%%%%%%%%%%%%%%%%%%%%%%%%%%%%%%%%%%%%%%%%%
%%%%%%%%%%%%%%%%%%%%%%%%%%%%%%%%%%%%%%%%%%%%%%%%%%%%%%%%%%%%%%%%%%%%%%%%%%%%%%%%%%
%%%%%%%%%%%%%%%%%%%%%%%%%%%%%%%%%%%%%%%%%%%%%%%%%%%%%%%%%%%%%%%%%%%%%%%%%%%%%%%%%%
%%%%%%%%%%%%%%%%%%%%%%%%%%%%%%%%%%%%%%%%%%%%%%%%%%%%%%%%%%%%%%%%%%%%%%%%%%%%%%%%%%

\end{document}